\documentclass[aps,pra,reprint, amsmath, amssymb,superscriptaddress]{revtex4-1}

\usepackage{graphicx}
\usepackage{textcomp} 
\usepackage{gensymb} 
\usepackage{dcolumn}
\usepackage{bm}
\usepackage{amsmath, amssymb}
\usepackage{dsfont}
\usepackage{float}
\usepackage{afterpage}
\usepackage{xcolor}
\usepackage[normalem]{ulem}
\usepackage{babel}
\usepackage{cancel}
\usepackage{hyperref}
\hypersetup{colorlinks, citecolor=blue, urlcolor=blue}
\usepackage{soul,xcolor}
\setstcolor{red}

\newcommand{\ket}[1]{\vert #1 \rangle}

\newcommand{\braket}[2]{\langle #1 \vert #2 \rangle}

\begin{document}

\title{Quantum Kernel Evaluation via Hong-Ou-Mandel Interference}

\author{C. Bowie}%
\affiliation{Centre for Engineered Quantum Systems, School of Mathematics and Physics, University of Queensland, QLD 4072 Australia}
\author{S. Shrapnel}%
\affiliation{Centre for Engineered Quantum Systems, School of Mathematics and Physics, University of Queensland, QLD 4072 Australia}
\author{M. J. Kewming}
\affiliation{School of Physics, Trinity College Dublin, College Green, Dublin 2, Ireland}

\date{\today}

\begin{abstract}
One of the fastest growing areas of interest in quantum computing is its use within machine learning methods, in particular through the application of quantum kernels.

Despite this large interest, there exist very few proposals for relevant physical platforms to evaluate quantum kernels. 
In this article, we propose and simulate a protocol capable of evaluating quantum kernels using Hong-Ou-Mandel (HOM) interference, an experimental technique that is widely accessible to optics researchers. Our proposal utilises the orthogonal temporal modes of a single photon, allowing one to encode multi-dimensional feature vectors. As a result, interfering two photons and using the detected coincidence counts, we can perform a direct measurement and binary classification. This physical platform confers an exponential quantum advantage also described theoretically in other works. 
We present a complete description of this method and perform a numerical experiment to demonstrate a sample application for binary classification of classical data.  

\medskip
\noindent{\it Keywords}: Quantum Optics, Machine Learning, Quantum Kernel Methods
\end{abstract}
\maketitle

\section{Introduction}
In recent years, there has been growing interest in the applications of machine learning to the physical sciences \cite{Carleo_machine_2019}.
One particular area of that has received considerable attention is quantum machine learning \cite{Carleo_machine_2019, wittek_quantum_2014, biamonte_quantum_2017, schuld_supervised_2018, ciliberto_quantum_2018, Schuld_circuit_2020, Cerezo_2021, Lloyd_2014, Rebentrost_quantum_2014, Lloyd_2016, Huang_2021, Havl_ek_2019, schuld_quantum_2019, kubler_quantum_2019, schuld_supervised_2021, Heyraud_noisey_2022, Liu_Differentialble_2018, Coyle_2020, wiebe_quantum_2016, Amin_quantum_2018}. 
In quantum machine learning, evaluating a quantum kernel is analogous to computing a classical kernel in classical machine learning. Both methods aim to measure the similarity or distance between data points in high-dimensional feature spaces without explicitly transforming the data. In classical machine learning, kernel methods use kernel functions to calculate inner products between data points in the transformed space. In quantum machine learning, classical data is encoded into quantum states, and the quantum kernel computes inner products between these quantum states to quantify their similarity or distance in the quantum feature space. Encoding classical data into a quantum system is equivalent to embedding the data into a quantum feature space \cite{Rebentrost_quantum_2014, chatterjee2016generalized, schuld_quantum_2019, Havl_ek_2019, kubler_quantum_2019, Schuld_circuit_2020, Lloyd_2020}, and measurement of the quantum system is equivalent to evaluating the kernel. This connection between classical and quantum kernels leverages the computational advantages of quantum computing for efficient inner product calculations, providing a powerful tool for quantum machine learning algorithms and applications in data analysis and pattern recognition.

Despite much success in the theoretical investigation of quantum kernels, there have been only a few experimental demonstrations. A seminal approach using nuclear spins \cite{Kusumoto_2021} encodes specific feature vectors into two unitary operators, which are consecutively applied to an initial state before the system is measured along its magnetic moment, providing kernel evaluation. Several applications using quantum optics have also been demonstrated. Two studies have encoded features into the dual-rail encoding of multiple photons, first demonstrated in \cite{Cai_entanglement_2015} and subsequently in \cite{Bartkiewicz_2020}. The entanglement between dual-rail encoded photons can be used to exploit the quantum advantage inherent in quantum kernels, namely the exponential speed-up in their evaluation \cite{lloyd2013quantum}.
A final example involves using the spectral modes of ultrafast radiofrequency pulses to classify and train labeled datasets \cite{Denis_photonic_2022}. The advantage of this methodology is that it makes use of the photon's spectral modes, decomposing these into an orthogonal eigenbasis for encoding higher-dimensional feature vectors. The latter results discussed above highlight the utility of photonics in quantum information. In fact, optically encoded quantum information presents itself in many machine learning examples \cite{Steinbrecher_2019, Killoran_continuous_2019, Taballione_2021, Chabaud_2021, Ghobadi_nonclassical_2021, Gan_2022}. The popularity of photonic quantum information stems from the photon's versatility given its numerous and highly controllable degrees of freedom \cite{Slussarenko_2019}. Despite the numerous encoding proposals that hinge on these increased degrees of freedom, the currently proposed physical platforms are limited to qubit-based models.
To this end, we propose a new method of evaluating quantum kernels using Hong-Ou-Mandel interference. As we will show, this method not only makes use of the entanglement necessary for the enhancement in quantum kernels \cite{lloyd2013quantum, Cai_entanglement_2015}, but also utilizes the higher-dimensional spectrum of single photons, which were exploited in \cite{Denis_photonic_2022}.

\section{Kernel Method Machine Learning}

We will begin by providing a brief pedagogical overview of kernel methods---for completeness see Ref.~\cite{hofmann_kernel_2008}.
To summarize Kernel methods succinctly, one starts by encoding data into some higher-dimensional feature space where classification of the data can be easier to analyse. 
However what makes this algorithm so useful is that is does not need to explicitly perform evaluations of the data in the feature space, but rather can be carried out using the kernel function that is defined on the domain of the original input data; this is commonly known as the \emph{Kernel trick}.

To understand this more precisely, let us describe a simple machine learning classification task. 
Suppose we have a data set $Y$---say of $N$ images---as a list of input vectors and labels, $Y \equiv \{(\vec{x}_{1}, y_{1}), (\vec{x}_{2}, y_{2}), \hdots (\vec{x}_{N},y_{N})\}$.
Here $y_{i}$ is the label---for example, cats and dogs---and $\vec{x}_{i}$ is the input data $\mathcal{X}$---for example the pixel colour values, or some other set of features.
The goal of our classification task is ultimately to classify an unknown data set $Y'$, that is a set of data with unknown labels.
One approach is to determine this complex relationship between inputs $\vec{x}_{i}$ and labels $y_{i}$, such that if it is presented with an unknown data set $Y'$, the algorithm will correctly classify this data.
This is typically the approach of Deep Neural Networks which employ a vast number of tunable parameters and non-linear functions to effectively replicate this relationship.

\begin{figure}
    \centering
    \includegraphics[width=\columnwidth]{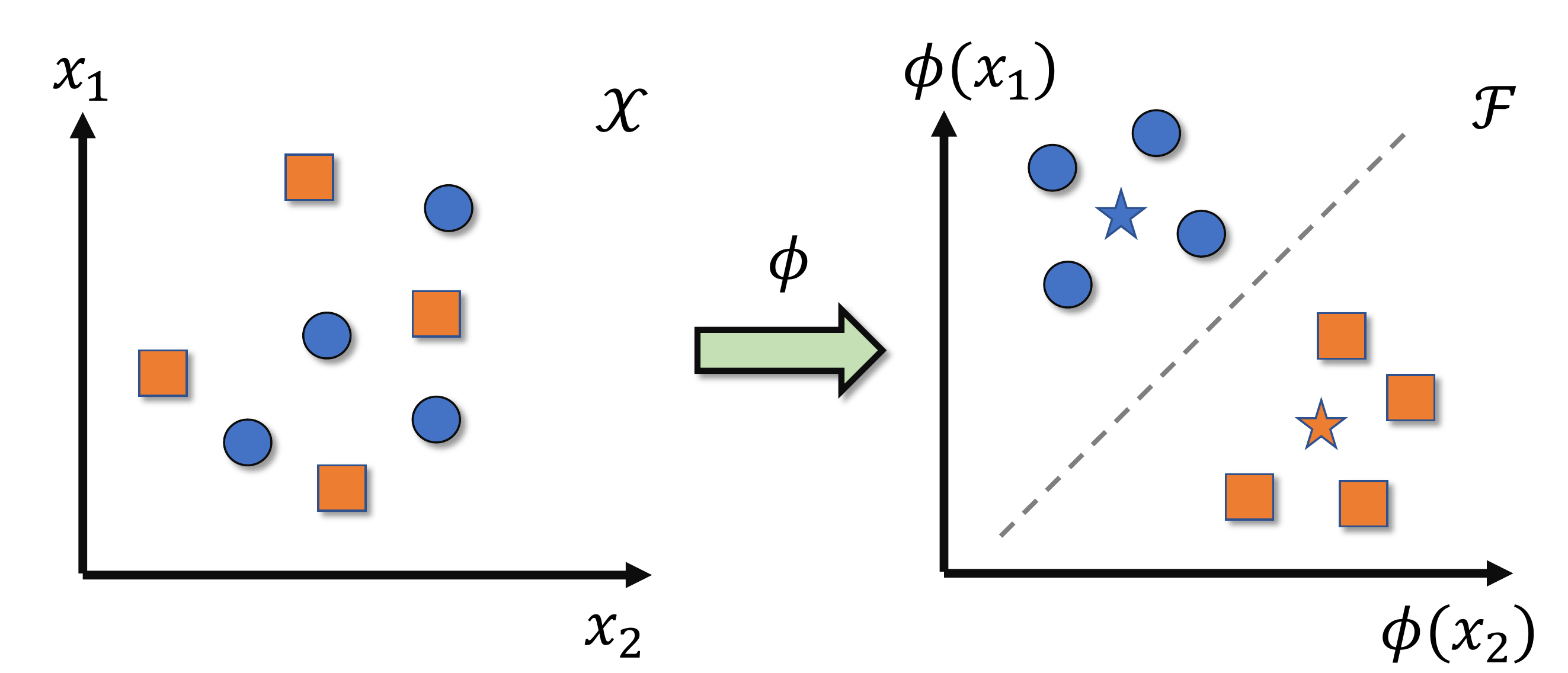}
    \caption{
    A visual representation of kernel machine learning, between two classes; blue circles and orange squares. (left) A depiction of the initial feature space of unprocessed data contained in the 2 dimension features space with features $x_1$ and $x_2$. (right) After transforming to the feature space $\mathcal{F}$ via $\phi$, the data can clearly be linearly separated into two classes (for illustrative purposes, we have limited this to a two dimensional transform, but this can be relaxed in general). In this depiction, the average of each class can be computed (coloured stars) and used as a discrepancy measured via MMD.}
    \label{kernel fig}
\end{figure}

Kernel methods on the other hand embed the data $\mathcal{X}$ in a higher-dimensional feature space $\mathcal{F}$ via the feature map $\phi{:\mathcal{X}\rightarrow \mathcal{F}}$.
In the feature space, the data is then classified mathematically by a suitable choice of distance measure along a decision boundary---for example, the inner product which we will consider going forward.
This concept is demonstrated visually in Fig.~(\ref{kernel fig}). 
In this sense, a kernel $k$ maps two inputs $\vec{x}$ and $\vec{x}'$---both of which are from the input space $\mathcal{X}$---to a distance measure $\mathbb{C}$ such that $k: \mathcal{X} \times \mathcal{X} \rightarrow \mathbb{C}$ \cite{hofmann_kernel_2008}.
Moreover, the feature map $\phi$ is related to the kernel mapping via the inner product of different feature vectors
\begin{equation}
\label{eq:kernel}
    k(\vec{x},\vec{x}') = \langle \phi(\vec{x}), \phi(\vec{x}') \rangle \,,
\end{equation}
which has an associated Gram matrix $K$,
\begin{equation}
    K_{m,n} = k(\vec{x}_{m},\vec{x}_{n}) \forall \vec{x} \in \mathcal{X}\,,
\end{equation}
that is positive semi-definite.
If the Gram matrix satisfies the condition
\begin{equation}
    \label{gpsd}
    \sum_{m,n}^M c_m c_n*k(\vec{x}_m,\vec{x}_n) \geq 0\,,
\end{equation}
for any $c_1 . . . c_M \in \mathbb{C}$, an associated feature mapping is guaranteed to exist and the function can be considered a kernel \cite{hofmann_kernel_2008}. 
With all this specified one can then classify a data point $\vec{x}$ in the feature space $\mathcal{F}$ according to some decision boundary $b$ that separates the classes as depicted in Fig.~(\ref{kernel fig})
\begin{equation}
    f(x) = \langle b, \phi(\vec{x})\rangle,
\end{equation}
where $f(x)$ is positive or negative indicating the binary classification.

Given that a kernel can be evaluated as the inner product between two feature vectors, there is a natural extension to quantum feature spaces \cite{schuld_quantum_2019}.
Suppose that we again have some input data $\vec{x}$, which is then encoded into a quantum state $\ket{\Phi(\vec{x})}$.
The quantum kernel will correspond to the overlap of this state with another
\begin{equation}
\label{eq:quantum_kerneal}
    k(\vec{x}, \vec{x}') =  \braket{\Phi(\vec{x})}{\Phi(\vec{x}')}\,.
\end{equation}
Furthermore, given that all the kernel outcomes can be mapped to probabilities, any inner product calculated using quantum states way will automatically satisfy the gram matrix conditions outlined in Eq.~(\ref{gpsd}).
This has provided ample motivation for the development of kernel based quantum machine learning. 

Evaluating the kernel requires that one can directly measure the overlap between two quantum states, which either requires quantum state tomography or other methods using quantum computers where one can directly parameterize the feature map $\phi$.

\section{Temporal encoding and Hong-Ou-Mandel Interference}

Here we will outline a proposal for quantum kernel evaluation using the temporal encoding of single photon Fock-states and Hong-Ou-Mandel interference. 
Creating higher dimensional quantum optical states can be achieved in many ways as detailed in the recent reviews on this topic \cite{Rubinsztein_Dunlop_2016, Erhard_2017, Flamini_2018, Shen_2019, Slussarenko_2019}.
Experimentally however, not all methods of encoding information in optics are made equal. For example, a natural orthonormal basis would be multi-photon states, however these can be difficult to measure and require photon number resolving detectors \cite{jonsson_evaluating_2019, fitch_photon-number_2003, lee_nearly_2015, provaznik_benchmarking_2020}.

A more common methodology which circumvents this approach is to encode into multiple photons in different paths, while simultaneously exploiting the two polarisation modes, thus creating numerous dual rail encoded qubits, which to date have been used to evaluate quantum kernels \cite{Cai_entanglement_2015}. However, the physical scaling and technological challenges associated with nuermous dual rail qubits can prevent higher dimensional feature spaces from being reached in practice.

We propose to sidestep these complications through the use of temporally/frequency encoded photons, a method which has received considerably less attention in the quantum kernel community \cite{Kewming_designing_2021, milburn_physics_2022}.
A continuous-mode single-photon state can be described as the coherent superposition of many spectral modes $\omega$
\begin{equation}
\label{eq:single_photon}
    \ket{1_{\Psi}} = \int d\omega \Psi(\omega) \hat{a}^{\dagger}(\omega)\ket{0}\,,
\end{equation}
where $\Psi(\omega)$ is the spectral density function that weights each mode, and $\hat{a}^{\dagger}(\omega)$ are the creation operators associated with each $\omega$.
Furthermore, we will make the quantum optics approximation whereby we assume that the spectral spread is much smaller than the carrier $\omega\ll \omega_{c}$ \cite{walls_quantum_2008}.
In this limit, the Fourier transform of the slowly varying spectral envelope corresponds to the temporal wave-packet, thus yielding the description of the single photon state defined in the time domain
\begin{equation}
    \ket{1_{\Psi}} = \int dt \Psi(t) \hat{a}^{\dagger}(t)\ket{0}\,.
\end{equation}
Here the temporal wave-packet satisfies the normalisation condition $\int dt \vert \Psi(t)\vert^{2} {=} 1$ and the bosonic field operators satisfy the commutation relation 
\begin{equation}
\label{eq:commuator}
    \left[\hat{a}(t), \hat{a}^{\dagger}(t')\right] = \delta(t-t')\,.
\end{equation}

Feature vectors can be encoded into the temporal modes of single photons provided a set of orthogonal temporal modes are chosen
\begin{equation}
    \Psi(\vec{x}, t) = \sum_{n=1}^N \alpha_n(\vec{x}) u_n(t) 
\end{equation}
where $\alpha_n(\vec{x})$ denotes a unit weight---encoding the information from the feature vector---of each mode and $\{u_n(t)\}$ a set of orthonormal temporal mode functions  \cite{milburn_physics_2022}. 
In our work, we will take the set $\{u_n(t)\}$ as the set of orthogonal Hermite Gaussian (HG) modes, noting that this choice is arbitrary could be replaced by any numerically orthogonal set of single-variable functions,
\begin{equation}
    u_{n}(t) = \frac{e^{-\frac{t^{2}}{2}}}{\pi^{1/4}\sqrt{2^{n}n!}}H_{n}(t) \,,
\end{equation}
where 
\begin{equation}
    H_{n}(t)=  (-1)^{n}e^{-t^{2}}\frac{d^{n}}{dt^{n}}e^{-t^{2}/2}\,,
\end{equation}
is the $n^{\mathrm{th}}$ Hermite polynomial and satisfies the orthogonality relation
\begin{equation}
\label{eq:orthogonality}
    \int d t u_{n}^{*}(t)u_{m}(t) = \delta_{nm}\,. 
\end{equation}
Furthermore, we will define the unit weight vector $\alpha_n$ simply as
\begin{equation}
\label{eq:weights}
    \alpha_n(\vec{x}) = w_{n} \phi_{n}(\vec{x})\,,
\end{equation}
 where $\phi_{n}(\vec{x}){\in }\mathbb{C}$ is the $n$th element of the feature vector of the input data $\vec{x}$, and $w_{n}$ is a free weight---commonly added in kernel methods to allow optimisation of the resulting kernel.
 Moreover the coefficients are normalized such that $\sum_{n}\vert \alpha_{n} \vert^{2}{=} 1$. 
 We are thus in a position to now define the proposed single photon encoding for an input vector $\vec{x}$ as
\begin{equation}
\label{eq:photon_encoding}
    \ket{\Phi(\vec{x}, t)} = \sum_{n=1} \int dt \alpha_{n}(\vec{x}) u_{n}(t) \hat{a}^{\dagger}(t) \ket{0}\,.
\end{equation}
In our proposed encoding scheme, the single photon corresponds to the information carrier, the HG temporal modes correspond to the orthogonal basis of the feature space, and the coefficients encode the data into this basis.

\begin{figure}
    \centering
    \includegraphics[width=\columnwidth]{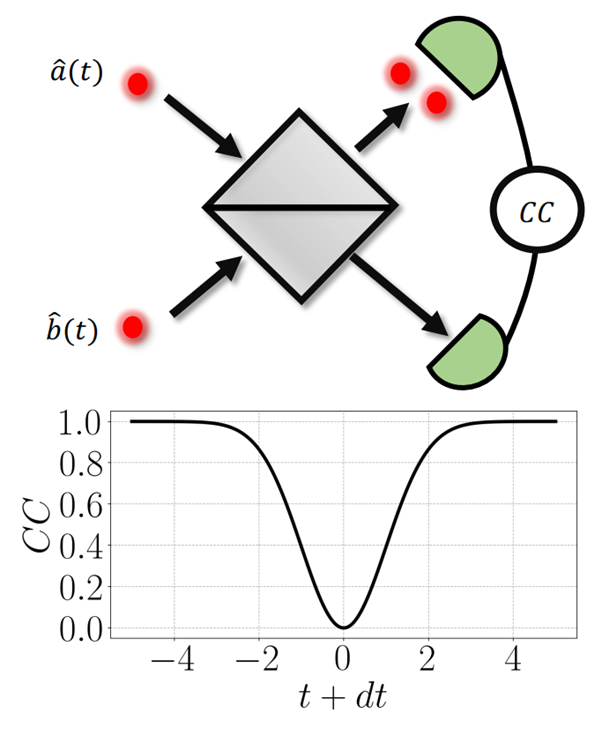}
    \caption{(Top) A depiction of HOM interference where two identical photons incident on a beam splitter simultaneously will interfere, causing them to bunch into pairs. This effect can be measured using the coincidence counts (${\rm CC}$) of the two respective detectors. (Below) Shows the HOM dip as a function of the relative time-delay $dt$ of the photons. As the time-delay between the two vanish, then the probability of measuring a ${\rm CC}$ also vanishes.}
    \label{fig:HOM_depiction}
\end{figure}

Now suppose we have two encoded feature vectors $\ket{\Phi(\vec{x},t)}$ and $\ket{\Phi(\vec{x}',t)}$ and we would like to measure the quantum kernel Eq.~(\ref{eq:quantum_kerneal}).
We can do this by interfering the two photons with each other on a 50:50 beam-splitter (BS).
This interference is intrinsically quantum mechanical and leads to `bunching' where both photons exit the same port and are detected together. This is known as Hong-Ou-Mandel (HOM) interference \cite{hong_measurement_1987}.
The probability of detecting the two photons simultaneously at either detector $P(2,0)$ or $P(0,2)$ is equal to the overlap of the two wave functions
\begin{equation}
\label{eq:P2}
    P(2,0) + P(0,2) = \vert \braket{\Phi(\vec{x}',t')}{\Phi(\vec{x},t)}\vert^{2}\,,
\end{equation}
which can be readily evaluated using the commutation relation Eq.~(\ref{eq:commuator}) and the orthogonality of the HG modes Eq.~(\ref{eq:orthogonality}) as
\begin{equation}
    \left|\braket{\Phi(\vec{x}',t')}{\Phi(\vec{x},t)}\right|^{2} = \left|\sum_{n}\alpha_{n}^{*}(\vec{x}')\alpha_{n}(\vec{x})\right|^{2}\,,
\end{equation}
which provides a direct evaluation of the quantum kernel. 

This quantum kernel can be observed directly by measuring the HOM dip, whereby two temporally synchronized detectors measure correlated photon detection events, otherwise known as coincidence counts (CC) \cite{hong_measurement_1987, bouchard_two-photon_2021}. This is visualized in Figure.~(\ref{fig:HOM_depiction}). 
The normalized CC (divided by the total number of counts) is equal to the probability of measuring a photon event at both detectors simultaneously and is therefore equivalent to 
\begin{equation}
\label{eq:CC}
    \mathrm{CC}(\vec{x}',\vec{x}) = 1-\left|\braket{\Phi(\vec{x}',t')}{\Phi(\vec{x},t)}\right|^{2}\,.
\end{equation}
This provides a clear experimental evaluation of the quantum kernel where the HOM interference will classify the similarity between feature vectors $\vec{x}$ and $\vec{x}'$.

Finally, there is an immediate parallel between our physical proposal and a theoretical proposal subject to a quantum advantage, discovered in \cite{lloyd2013quantum} and tested in \cite{Cai_entanglement_2015}. Firstly, we are both working towards solving the same task, namely supervised cluster assignment. In the interest of brevity, we will use our abbreviated notation in Eq.~(\ref{eq:single_photon}), where $\ket{\Phi(\vec{x}', t)}{\rightarrow }\ket{1_{\Phi'}}$ describes the encoded features in orthogonal temporal modes. We therefore initially begin with two photons, encoded into different paths corresponding to the initial state:
\begin{equation}
\ket{\psi} = \ket{1_{\Phi'}}\ket{1_{\Phi}}.
\end{equation}
The two photons are then interfered on a single beam splitter as described above, leading to the transformation:
\begin{equation}
\label{eq:psi2}
\ket{\psi} \rightarrow \frac{1}{2}\left(\ket{2_{\vec{u},\vec{v}}}\ket{0} - \ket{0}\ket{2_{\vec{u},\vec{v}}} + \ket{1_{\vec{u}}}\ket{1_{\vec{v}}} - \ket{1_{\vec{v}}}\ket{1_{\vec{u}}}\right),
\end{equation}
where the notation indicates that the two photons are in the same path mode $\ket{2_{\Phi',\Phi}}$. As we have shown above, the probability of measuring two qubits (in either port) is given by Eq.~(\ref{eq:P2}), and measuring one qubit in both ports (a coincidence count) is given by Eq.~(\ref{eq:CC}).
This proposed scheme makes use of the same quantum resources required in \cite{lloyd2013quantum}, whereby a larger distance between $\Phi(\vec{x},t)$ and $\Phi(\vec{x}',t)$ is represented in the entanglement of the state given by Eq.~(\ref{eq:psi2}). Rather than relying on an empty ancilla as an indicator function, we make use of both photons simultaneously as information carriers. By utilizing entanglement as a resource in the same manner as \cite{lloyd2013quantum}, we conclude that our proposed physical platform also offers the same available quantum advantage over their classical counterparts.

\section{Application example: Numerical Experiment}
We have so far described how one can use the CC of temporally encoded photons to evaluate a quantum kernel. Now we will demonstrate an example application for a binary classification task using generated data. The model for machine learning and classification in this case is a maximum mean discrepancy (MMD) model via kernel mean embedding (KME), with training of the model performed via stochastic gradient descent (SGD). The applications of this physical platform are not limited to any aspect of this model, it can be used for any machine learning algorithm that requires kernel evaluation.

\subsection{Kernel Implementation Model: Maximum Mean Discrepancy }
This subsection outlines the use of MMD for kernel machine learning and specifically the implementation for our proposed physical platform. Suppose we have two classifications, both of which are characterized by the probability distributions over the input data $P(\vec{x})$ and $Q(\vec{x})$ respectively. 
Then given a kernel function ${k: \mathcal{X} \times \mathcal{X} \rightarrow \mathbb{C}}$ with associated mapping ${\phi:\mathcal{X}\rightarrow \mathcal{F}}$, we can define group mean $\mu_{P}$ over the classification group $P(X)$ as
\begin{align}
    \mu_{P} &= \int_{\mathcal{X_{P}}} k(\cdot, \vec{x}) d P(\vec{x})= \int_\mathcal{X_{P}} \phi(\vec{x})d P(\vec{x}) \,,
    \label{RepGroupMean}
\end{align}
where $\mathcal{X}_{P}$ indicates that we are only averaging over feature vectors of the $P$ class.
Our encoding scheme described above permits the immediate definition of the quantum feature mean \cite{kubler_quantum_2019}
\begin{equation}
\label{eq:mean_embedding}
    \ket{\mu_{P}} = \frac{1}{N_{P}}\int_{\mathcal{X}_{P}} \ket{\Phi(\vec{x}, t)}d P(\vec{x})\,,
\end{equation}
where $N_{P}$ is a normalisation constant. 
Therefore, we can create a quantum feature mean by simply averaging over all the state vectors corresponding to a quantum feature map.

We now define the maximum mean discrepancy (MMD) which corresponds to the absolute difference measure between any two distributions $P$ and $Q$, and is defined as the absolute difference between KMEs, 
\begin{align}
    \label{MMD}
    \mathrm{MMD} (P, Q) &= || \mu_{P} - \mu_{Q} ||^2\,, \\
    & = \langle \mu_{P}, \mu_{P} \rangle + \langle \mu_{
    Q}, \mu_{Q} \rangle - 2 \langle \mu_{P}, \mu_{Q} \rangle\,. 
\end{align}
What's more, if we now combine our quantum feature map using our single photon encoding Eq.~(\ref{eq:photon_encoding}) as well as our quantum feature mean Eq.~(\ref{eq:mean_embedding}), then we can evaluate the quantum MMD using our HOM interferometer where one photon encodes the mean $\mu_{P}$ and the other $\mu_{Q}$, which yields
\begin{equation}
    \mathrm{MMD} (P, Q) = 2 \mathrm{CC}(P,Q)\,,
    \label{MMDCC}
\end{equation}
where $\mathrm{CC}(P,Q) = 1 - \vert \braket{\mu_{P}}{\mu_{Q}}\vert^{2}$ given by Eq.~(\ref{eq:CC}).
We can therefore measure the MMD using a single evaluation of a HOM interferometer and the kernel function associated with feature mapping $\phi$. 
Moreover, using the free-weights $w_{n}$ in Eq.~(\ref{eq:weights}), one can optimize the MMD, yielding an optimal class separation in feature space $\mathcal{F}$. 
Crucially, the MMD model provides a cost function to be optimized in a given feature space allowing for an implicit separating hyperplane between classification groups. 
Once optimized, the mapping $\phi$ could be used to classify an unseen data point $\vec{x}'$, using a quantum HOM classifier by 
\begin{align}
    \mathrm{Class}(\vec{x})  &= \frac{1}{2}\left({\rm MMD}(\vec{x}, Q) - {\rm MMD}(\vec{x},P)\right) \,,\\
    &=\left|\braket{\Phi(\vec{x},t)}{\mu_{Q}}\right|^{2} -\left|\braket{\Phi(\vec{x},t)}{\mu_{P}}\right|^{2} \,,
    \label{eq:mmdclass}
\end{align}
where a positive evaluation of this class function will place $\vec{x}$ in class $Q$, and negative will place it in class $P$. 
If the feature map has been optimized such that the means are almost orthogonal ( $\braket{\mu_{P}}{\mu_{Q}} \sim 0$), then classification can be accurately approximated with a single feature mean: If, for example, $\vec{x}$ belongs to $P$, then the overlap with $\mu_{P}$ will be high, and low with $\mu_{Q}$, thus we would expect to see a HOM dip in the former, and not in the latter. 
To minimise the use of resources, we note that one does not need to compare $\vec{x}$ to both $\mu_{P}$ and $\mu_{Q}$, but rather just to one--- say $\mu_{Q}$---provided that the dip has been calibrated to ${\rm MMD}(P,Q)$. 
In this setting, one could set the decision boundary to be equal to 
\begin{equation}
    {\rm Class}(\vec{x}) \rightarrow 
    \begin{cases}
    Q & {\rm if} \left|\braket{\Phi(\vec{x},t)}{\mu_{Q}}\right|^{2} \geq \frac{1}{2} - \frac{1}{2}\left|\braket{\mu_{P}}{\mu_{Q}}\right|^{2}\\
    P &{\rm if} \left|\braket{\Phi(\vec{x},t)}{\mu_{Q}}\right|^{2} < \frac{1}{2} - \frac{1}{2}\left|\braket{\mu_{P}}{\mu_{Q}}\right|^{2}
    \end{cases}\,.
\end{equation}
In a high efficiency experiment (low loss and noise, high quantum efficiency) with highly orthogonal encoding ($\braket{\mu_{P}}{\mu_{Q}} \sim 0$),  this classification of an unknown $\vec{x}$ could be measured with very few single photon measurements as the expected coincidence count would be highly correlated or anti-correlated with the single comparison point $\mu_{Q}$. 
This ensures that very few photons are required to perform this classification, thus minimising the experimental overhead.

\subsection{Optimisation Training: Stochastic Gradient Descent (SGD)}
The MMD model for kernel implementation defines the criteria used to determine the ability of the model and the process of classification, but does not specify the optimisation algorithm used to train the model. 
For this example, we choose to implement a SGD optimisation algorithm, however one is free to choose whatever algorithm they see fit for this purpose. 

The parameter to be optimised are the free weights, given by the vector $\vec{w}$, which were introduced in Eq.~(\ref{eq:weights}). 
We use stochastic gradient descent (SGD) with MMD employed as the cost function to be optimized. This ensures that the weights are optimized with respect to the means, and most likely to be maximally orthogonal in the feature Hilbert space.
Moreover, we use SGD rather than normal gradient descent to ensure the optimisation does halt in any local minima.
As such the weights are updated at each iteration $i$ according to the difference equation
\begin{equation}
    \vec{w}_{i+1} = \vec{w}_{i}+ L_{i} \nabla_{\vec{w}} {\rm MMD}(P,Q)\vert_{\vec{w}=\vec{w}_{i}} + \vec{\epsilon}_{i} \,,
    \label{SGD}
\end{equation}
where the subscript $i$ indicates the iteration---not to be confused with the element $n$ in Eq.~(\ref{eq:weights})---and $\nabla_{\vec{w}} {\rm MMD}(P,Q)$ is the numerical derivative of the cost function evaluated at $\vec{w}_{i}$---MMD evaluation as given by Eq.~(\ref{MMD})---with respect to the weights. $L_{i}$ is then the learning rate at the $i$th iteration, and $\vec{\epsilon}_{i}$ is a stochastic random variable we add at each time step.

\subsection{Application and Results}
\begin{figure}
    \centering
    \includegraphics[width=\columnwidth]{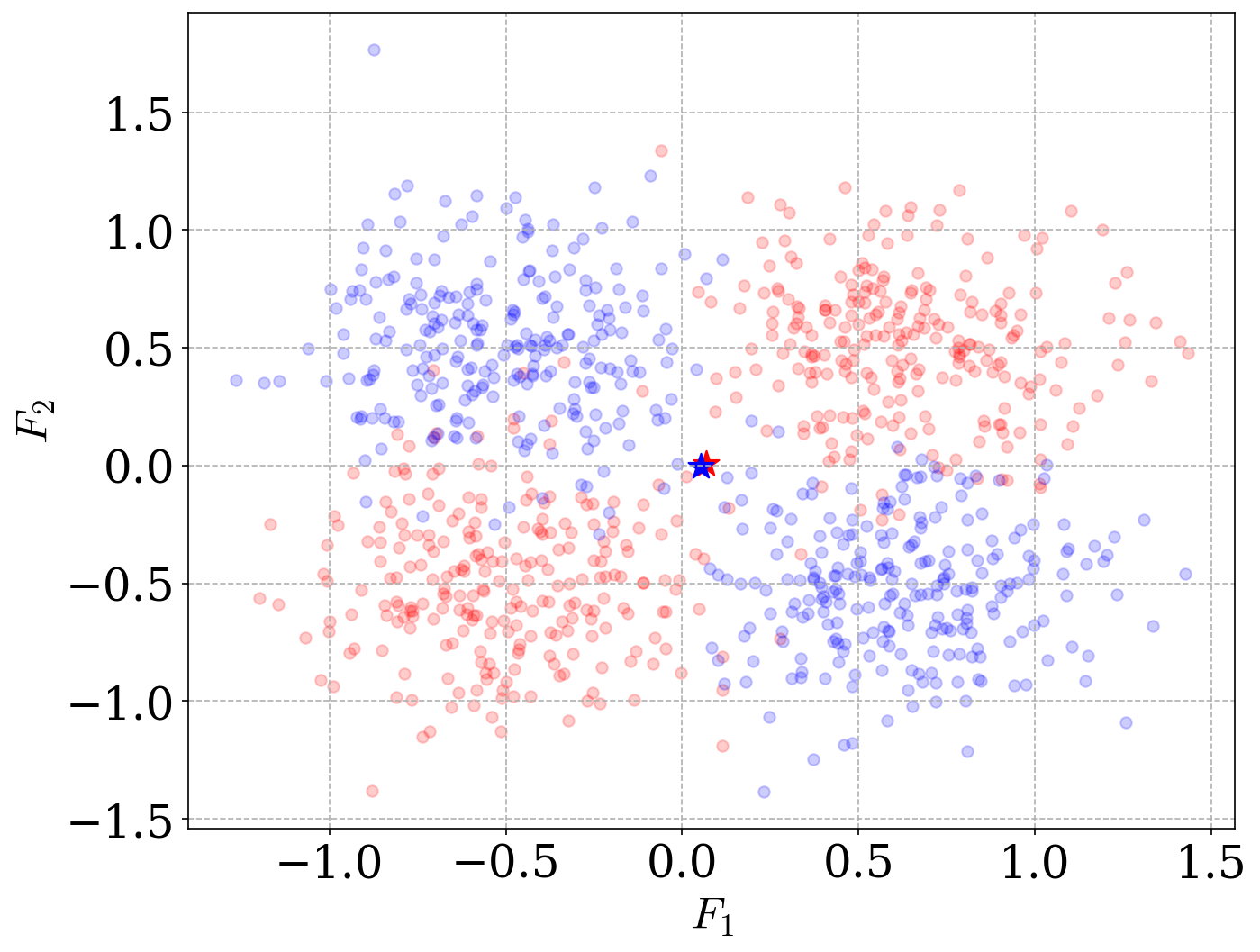}
    \caption{Scatter plot of the example data showing two classes (red) and (blue) which are clearly distinguishable, but have overlapping means and no hyperplane that separates them in two dimensions. Here $F_{1}$ and $F_{2}$ are two features and the mean data point for each classification group has noted by the symbol $(\star)$. 
    Mapping the means to feature space using the transformation in Eq.~(\ref{eq:mapping}) is insufficient to induce the needed hyperplane, separation only occurs in the feature space after optimisation of the free weight parameters } 
    \label{fig:example}
\end{figure}

To simulate the expected results of this model, a numerical experiment is performed. 
We initially generate a two dimensional (two features $F_{1}$ and $F_{2}$) training and test data set using scikit-learns' ``make blobs" function with parameters such that it would be separable in a polynomial feature space, but not be linearly separable \cite{scikit-learn}. The use of two feature is for visual demonstration purposes, and the model is applicable to higher dimensional data. 
Here the data corresponds to two classes blue $P$ and red $Q$.
The training data set is depicted in Fig.~(\ref{fig:example}).
From the training set, the group mean from each classification is determined by initially mapping each data vector to the feature space given by the polynomial kernel of degree two
\begin{equation}
\label{eq:mapping}
    \phi:(F_{1}, F_{2}) \rightarrow (F_{1}^{2}, F_{2}^{2}, F_{1}F_{2})\,,
\end{equation}
and then taking average values within each classification group. 
After this transformation, we can also add trainable weights $w_{n}$ as we introduced in Eq.~(\ref{eq:weights}), and then normalize them as suitable coefficients for the quantum encoding Eq.~(\ref{eq:photon_encoding}). The weights are then optimised using the SGD algorithm outlined in Eq. (\ref{SGD}). For the interested reader, a more detailed explanation of the training process and associated hyper-parameters can be found in Appendix \ref{sec:Appendix}.

\begin{figure}
    \centering
    \includegraphics[width=\columnwidth]{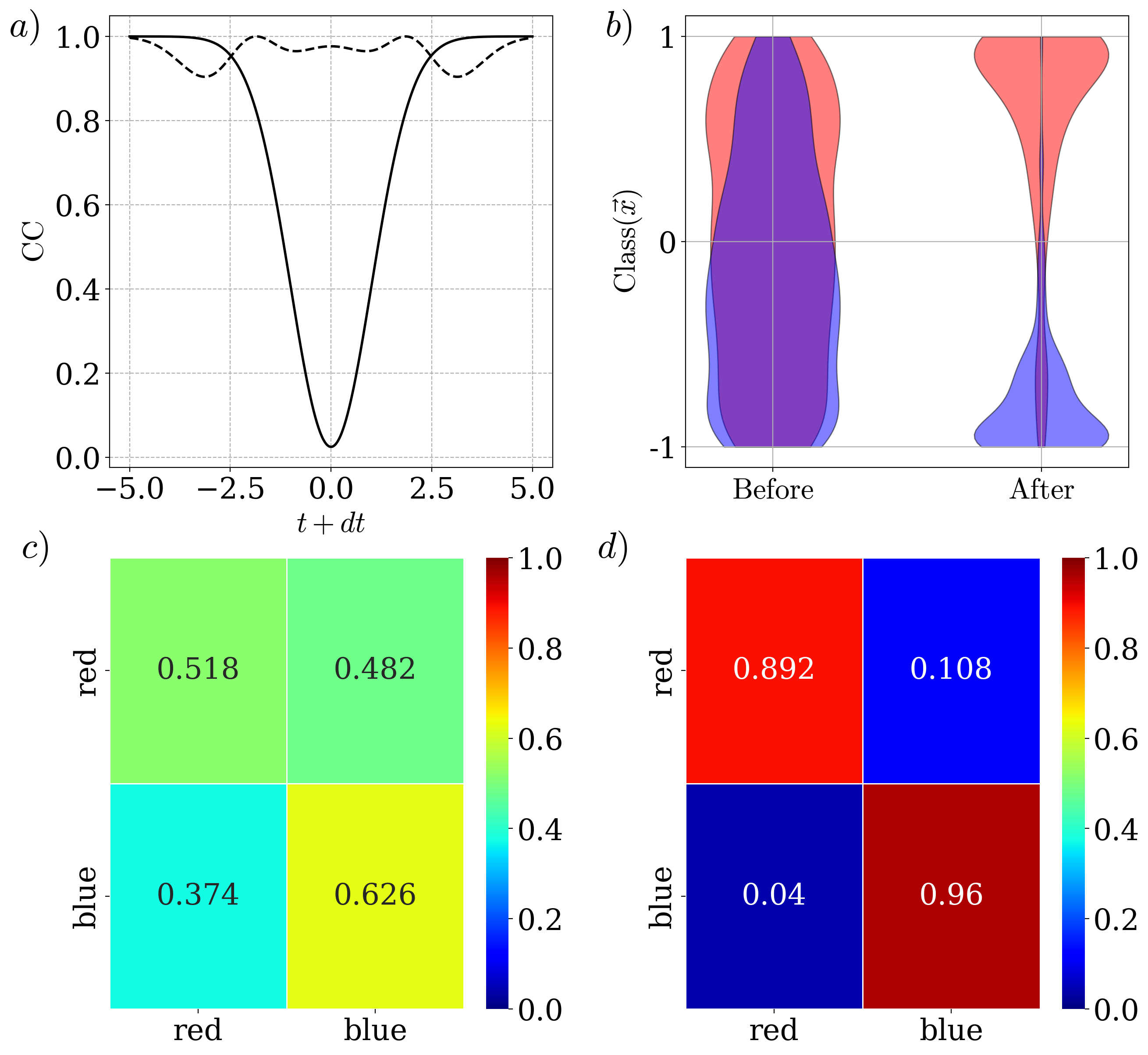}
    \caption{a) A depiction of the HOM dip of the two means created by evaluating the two means $\mu_{P}$ and $\mu_{Q}$ before (solid black line) showing a dip indicating significant overlap, and after training (black dashed line) showing no dip indicating a large separation of two means in the feature Hilbert space. b) Depicts two violin plots showing the classification of unseen data before and after using the quantum HOM classifier in Eq.~(\ref{eq:mmdclass}). 
    The widths of the violin plots are arbitrary but are proportional to the probability density of the classified distributions. After training the number of misclassifications by the HOM classifier is low. 
    Figure c) and d) Show the confusion matrix before and after training, where the diagonal (off diagonal) elements correspond to correct (incorrect) classifications.  }
    \label{poly results}
\end{figure}

The optimisation process maximises the MMD value of the two feature mean encoded photons $\mu_{P}$ and $\mu_{Q}$, minimising the probability of a HOM interference measurement registering a coincidence count. This process is demonstrated in 
Fig.~(\ref{poly results}a) where we plot the HOM dip calculated before and after the weights have been optimized between the two means $\mu_{P}$ and $\mu_{Q}$. At the center of the dip, where the time-delay between the photons is zero $dt=0$, the means are initially very similar, resulting in no discernible HOM dip---$\braket{\mu_{P}}{\mu_{Q}} \sim 1$---but after training are maximally orthogonal thus generate a large dip---$\braket{\mu_{P}}{\mu_{Q}} \sim 0$. Moreover, when the training is complete, we can use our encoded photons to classify unseen data accurately via Eq.~(\ref{eq:mmdclass}) which is depicted in the Violin plots in Fig.~(\ref{poly results}b) and also shown in the before and after training confusion matrices Fig.~(\ref{poly results}c) and Fig.~(\ref{poly results}d). 
The confusion matrix is a machine learning tool that demonstrates the percentage of correct/incorrect classifications. The vertical axis denotes the calculated classification as given by the machine HOM classifier, and the horizontal axis denotes the true classification.
We clearly see that after training we are able to classify the data precisely using our HOM quantum classifier.

\section{Discussion}
In this article we proposed a practical experimental methodology for evaluating quantum kernels using HOM interference of temporal encoded photons. 
We showed that the kernel is directly related to the conicidence counts measured via two photodetectors. 
We further showed that by using a quantum KME, one could also perform MMD for classification, which we demonstrated using a very simple example. 
Given that we have optimized the MMD between the two classes, then it is quite reasonable to assume that the number of experimental runs required to classify the data is minimized---since quantum classification via quantum kernel methods necessarily requires estimating the probability given by Eq.~(\ref{eq:quantum_kerneal}).
Finally, this concept offers another experimentally feasible methodology for evaluating quantum kernels in the growing application of photonic quantum machine learning.

\section{Acknowledgements}
We would like to acknowledge the useful conversations with Gerard Milburn and Sahar Basiri-Esfahani. 
This work was funded by the Australian Research Council
Centre of Excellence for Engineered Quantum Systems
(Project No. CE170100009).
MJK acknowledges the financial support from a Marie Sk\l odwoska-Curie Fellowship (Grant No. 101065974).
All code used to simulate all presented results is available in a devoted github repository \cite{cassiejayne_optical-kernel-simulation_2022}

\appendix

\section{\label{sec:Appendix}Details of numerical example}

\begin{figure}
    \centering
    \includegraphics[width=\columnwidth]{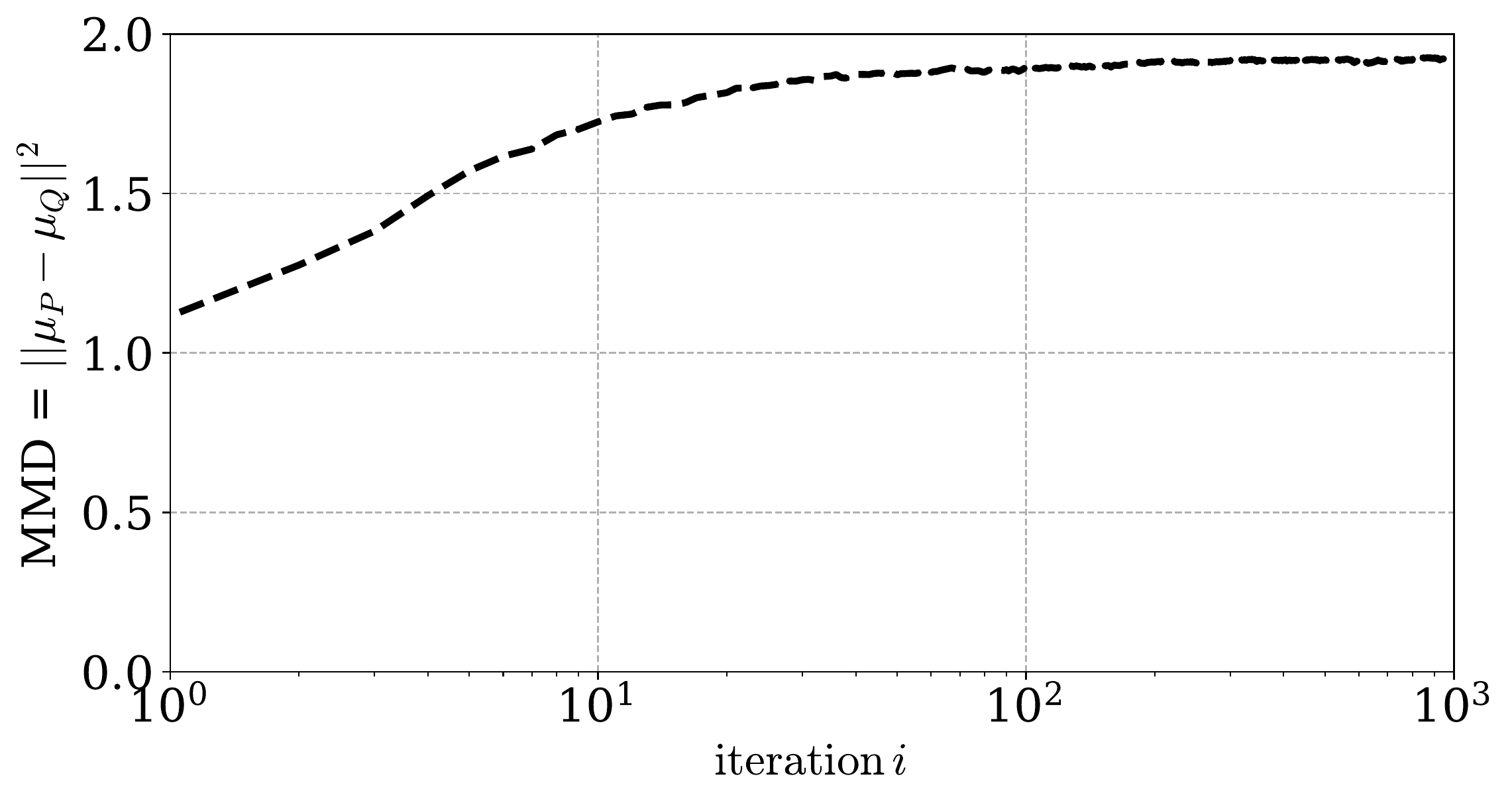}
    \caption{A plot of the average cost function (MMD value) over 1000 trials with regards to the epoch. The average initial parameters are randomly chosen, hence the non-zero overlap. For the first few iterations there is very little change, before the the algorithm begins to maximise the cost function. }
    \label{fig:Train}
\end{figure}

 Training was performed on 1000 data points with two features created using scikit's ``make blobs" function. Four blobs were created with different centering, and were grouped into two as demonstrated in Fig. \ref{fig:example}. This process was repeated to create test data, using the same centers to represent the same distribution, and the same number of data points. All simulations were run on a conventional desktop computer. 

Using the Maximum Mean Discrepancy (MMD) method, most of the analysis was conducted on the mean data value of each data set in the transformed feature space. Each data point was transformed according to the relevant mapping $\phi$, and then the data points for each classification group were grouped and averaged. At this stage, the averages were normalised for applicability to quantum encoding. 

The cost function to be optimised is the MMD as defined in Eq.~(\ref{MMD}). In realistic application, the MMD would be evaluated using Eq.~(\ref{MMDCC}). For the purposes of simulation, the MMD is evaluated mathematically as in Eq.~(\ref{MMD}). All notes in this section apply to the model explored in section five. 

The training is performed using a Stochastic Gradient Descent (SGD) algorithm. This is an iterative process that updates the free weights using the gradient of the MMD function with respect to the last iterative weight change. This process is described in Eq.~(\ref{SGD}). The training process is often subject to a number of hyper parameters. In our case, the hyper parameters employed are dependant on the current cost function, as the cost function is to be maximised, and bound by a maximum value of two. The training runs through 1000 iterations and then as the learning rate $L$ and the noise $\epsilon$ vary depending on the measure of the cost function. These are scheduled according to the the following table:
\begin{center}
\begin{tabular}{ | m{3cm} | m{5.2cm}|  } 
  \hline
  ${\rm Cost} < 1.8$& $L=0.1$, $\epsilon=\mathcal{N}(0, 0.5)$ \\ 
  \hline
  $1.8 <{\rm Cost} < 1.9$& $L=0.01$, $\epsilon=\mathcal{N}(0, 0.05)$ \\ 
  \hline
  $1.9< {\rm Cost}$& $L=0.001$, $\epsilon=\mathcal{N}(0, 0.05)$ \\ 
  \hline
\end{tabular}
\end{center}
where $\mathcal{N}(0, \sigma)$ corresponds to a normal distribution with mean 0, and standard deviation $\sigma$.

\bibliography{okp}
\bibliographystyle{apsrev4-1}

\end{document}